\documentstyle[12pt]{article}
\begin{document}
\def\singlespacing{\baselineskip=12pt}
\def\doublespacing{\baselineskip=24pt}
\def\d{\hbox{\bf d}}
\def\delt{{\bf \delta}}
\singlespacing
\noindent
\pagestyle{empty}
\today \\
\def\bdry{{\partial {\cal M}}}
\def\Qhat{{\widehat Q}}
\def\ens{{\cal E}}
\def\zbar{\overline}
\bigskip
\bigskip
\doublespacing
\font\mainh=cmbx10 scaled \magstep1
\baselineskip=24.9truept plus 2pt
\def\vphi{{\mbox{\boldmath $\phi$}}}
\def\J{{\bf J}}
\def\vx{{\bf x}}
\def\frac#1#2{{\textstyle {#1\over #2}}}
\medskip
\begin{center}
\begin{large}
{\mainh NICOLAI MAPS FOR QUANTUM COSMOLOGY } \\
\end{large}
\bigskip
\bigskip
R. GRAHAM\\
\smallskip
Universit\"at Gesamthochschule Essen\\
W-4300 Essen 1\\
Germany\\
\bigskip
\medskip
H. LUCKOCK\\
\smallskip
School of Mathematics and Statistics\\
University of Sydney\\
NSW 2006\\
Australia\\
\end{center}
\bigskip
\bigskip
\bigskip
\leftline{\bf Abstract}\smallskip\noindent

We construct Nicolai maps for $N=2$ supersymmetric extensions of minisuperspace
models. It is shown that Nicolai maps exist for only a very restricted set of
states. In the models considered these are the two states corresponding to the
empty and the filled fermion sectors. The form of the Nicolai maps in these
sectors is given explicitly, and it is shown that they have a natural
stochastic
interpretation. This result also suggests a probabilistic interpretation of the
wave function.

\newpage
\pagestyle{plain}
\pagenumbering{arabic}

\leftline{\mainh Introduction}
\smallskip
One of the most problematic features of quantum cosmology is that the
Hamiltonian
constraint is quadratic in the momenta. Two resulting difficulties
are that of intepreting the wave function, and that of ascertaining the
initial conditions for the Universe.

It has recently been discovered that the minisuperspace Wheeler-DeWitt
equations for certain Bianchi models admit simple supersymmetric extensions,
for which the Hamiltonian operator can be represented as the anticommutator
of two first-order operators \cite{DH,Gr1,Gr2}. This means
that the problematic second-order Hamiltonian constraint can be replaced by two
first-order constraints. The possible consequences of this result have only
begun
to be explored. It is conceivable that both of the problems mentioned above
could be
disposed of using supersymmetry.

One of the remarkable features of supersymmetric theories is that they
admit Nicolai maps, which means that they can be transformed into
non-interacting bosonic theories \cite{Nic}. When realistic boundary conditions
are
imposed, however, there are limits to the applicability of this result;
only certain quantum states can actually be generated by Nicolai maps
\cite{L1}.

In supersymmetric quantum cosmology, the wave function of the Universe
consists of a number of linearly independent components, corresponding
to quantum states satisfying different boundary conditions. A question of
considerable current interest is the choice of the boundary condition.
The Hartle-Hawking no-boundary proposal is one possible prescription \cite{HH}.
Here we investigate, in the context of $N=2$ models, the consequences of
requiring the existence
of a Nicolai map. We find that this requirement allows just two states;
one each in the empty and filled fermion sectors. Of these two states, at most
one can be normalisable. It is interesting to note that the same two
states are singled out by Lorentz invariance in some mini-superspace models
with extended $N=4$ supersymmetry obtained from $N=1$
supergravity by dimensional reduction \cite{DHO,Gr4,D2}.

Because Nicolai maps are naturally interpreted as stochastic processes,
this approach suggests an intepretation of the wave
function (in an appropriate representation) as a probability density
whose integral is conserved in Euclidean time.
\bigskip

\leftline{\mainh Euclidean Quantum Mechanics}
\smallskip
It is well-known that the ground state of a quantum mechanical system can be
found by rotating to Euclidean time $\tau=i t$ and performing a functional
integral over
paths whose actions vanish in the far past. This is the motivation for the
Hartle-Hawking proposal for the ground state of quantum cosmology \cite{HH}.

Since the initial conditions are set in the far past, it follows that the wave
function
obtained by this procedure is invariant with respect to Euclidean time
translation.
In other words, this procedure ensures that the wave function is annihilated by
the
Hamiltonian operator for the Euclidean theory. But it is easily seen that the
Hamiltonian operator is exactly the same in the Euclidean formulation as in the
usual Lorentz formulation, so that the wave function is a zero-energy state
of the original ``Lorentzian" theory.

Although the Hamiltonians for the two theories are represented by the same
differential operator, the momenta are not.
In the usual Lorentzian theory, the momentum operators are defined as
$\widehat p_\mu=-i\hbar \partial/\partial q^\mu$ so that
\begin{equation}
\widehat p_\mu \exp {i\over\hbar}S ={\partial S\over\partial q^\mu} \exp
{i\over\hbar}S.
\end{equation}
In the Euclidean theory, however, the momentum operators are defined as
$\widehat\pi_\mu=-i\widehat p_\mu = -\hbar\partial/\partial q^\mu$
so that
\begin{equation}
\widehat\pi_\mu \exp  -{1\over\hbar}S_E= {\partial S_E\over\partial q^\mu}
\exp -{1\over\hbar}S_E
\end{equation}
where $S_E$ is the Euclideanised action.

The Euclidean momentum operators $\widehat\pi_\mu$ are
not hermitian, but in fact this is not necessary. In standard ``Lorentzian"
quantum mechanics, observables must be represented by hermitian operators so
that
they have real expectation values. In the Euclidean case, however, all wave
functions
are real and so any \it real \rm operator will have a real expectation value.
Consequently, observables in Euclidean quantum mechanics are represented by
real (rather than hermitian) operators.

The reality of the Euclidean wave function can be seen most directly from its
path-integral representation,
\begin{equation}
\Psi(q_1,\tau_1) =\int_{q(\tau_1)=q_1} d[q(\tau)] \exp -{1\over\hbar}
S_E[q(\tau)]
\end{equation}
where the integral is taken over all histories for which $q(t')=q'$ and which
satisfy
some condition in the past which specifies the state.
This suggests that $\Psi$ could be interpreted simply as a probability, rather
than any
kind of quantum amplitude. This interpretation appears even more natural when
the model
is supersymmetric, since then the probabilities automatically add up to one
\cite{zum}.
\bigskip

\leftline{\mainh Supersymmetry and Nicolai Maps}
\smallskip
{}From our point of view, Euclidean quantum mechanics is characterised by
another
important feature; if the theory is supersymmetric then generally only the
Euclidean version will admit a Nicolai map.

Indeed, the original proof of Nicolai's theorem was given in Euclidean
spacetime \cite{Nic}. It is sometimes possible to rotate Nicolai maps from
Euclidean
to Lorentzian spacetime, but this generally involves a complex rotation of the
bosonic
variables $q^\mu$ as well. If these variables label points in a real
configuration
space, such as the components of the metric on a Riemannian manifold for
example,
it may be unreasonable to treat these as complex variables. In such cases we
must
therefore stay with the Euclidean formulation if we want to construct Nicolai
maps.

Nicolai maps in supersymmetric Euclidean theories are naturally interpreted as
stochastic differential equations which describe the evolution of the system in
Euclidean time. From a physical point of view, this evolution is of little
interest
in itself; however the stationary state which is finally reached is of
considerable
interest, since it represents the ground state of both the Euclidean and the
Lorentzian theory.

It has recently been found that a number of minisuperspace formulations of
quantum cosmology admit simple supersymmetric extensions. Because the action
for
quantum cosmology is invariant under local reparametrisations of the time
coordinate,
any supersymmetry in the quantum theory should be local with respect to time.

A Nicolai map for a locally supersymmetric theory can be obtained by
integrating
out the remaining fermionic fields, after imposing some gauge condition for all
local
symmetries so that there is a contribution from only one member of each class
of
physically equivalent configurations. In particular, in the case of locally
supersymmetric quantum mechanics there are fermionic gauge fields \cite{BDZDH}
and one can choose a gauge in which these vanish everywhere. (This is
possible here because the gauge field and the parameter of infinitesimal
fermionic
supersymmetry transformations have the same number of components in this case.
Note, however, that it would not be possible to eliminate the gauge field in
higher dimensional theories such as superstrings or inhomogeneous
supergravity.)

To avoid introducing ghosts, it is also necessary to fix the bosonic
reparametrisation
symmetry. This is done most simply be requiring that $\dot N=0$ \cite{Hal},
after
which we find that the theory has reverted to its original globally invariant
form.
The gauge fields have gone and the lapse function is constant; the only
supersymmetry
transformations which preserve these conditions are those parametrised by
constants
-- i.e. rigid transformations. One is thus drawn to the conclusion that,
in the case of supersymmetric quantum mechanics, the local theory is just the
global
theory in disguise.

Nicolai maps can therefore be obtained directly from the rigid theory, without
bothering
to consider the locally invariant version. In particular, this will apply to
supersymmetric descriptions of homogeneous cosmologies.
\bigbreak

\leftline{\mainh N=2 Supersymmetric Non-Linear $\sigma$-Model}\nobreak
\smallskip\nobreak
The one-dimensional N=2 supersymmetric non-linear $\sigma$-model provides a
supersymmetric description of a particle moving in a curved configuration
space \cite{SSQM}.
Originally suggested by Witten as a toy model for understanding the quantum
mechanics of general supersymmetric theories, it applies very neatly to
the supersymmetric models of minisuperspace quantum cosmology which have
recently
attracted attention.

This model is described by a Lagrangian of the form
\begin{equation}
L\equiv
\frac12 G_{\mu\nu}{dq^\mu\over d\tau} {dq^\nu\over d\tau} +\frac12
G^{\mu\nu}\Phi_{,\mu}\Phi_{,\nu}
+\frac12 (\psi^*_\mu {D\psi^\mu\over d\tau} -{D\psi^*_\mu\over d\tau} \psi^\mu)
+\Phi_{;\mu\nu}\psi^{*\mu}\psi^\nu +{1\over 4} R_{\mu\nu\rho\sigma}
\psi^{*\mu}\psi^{*\nu}\psi^\rho\psi^\sigma  \label{eq:Lagrangian}
\end{equation}
where $G_{\mu\nu}$ is the metric of the configuration space in which the
particle moves,
and $\Phi$ is a potential on this space. The covariant derivatives of the
fermion fields
are defined as
\begin{equation}
{D\psi^\mu\over d\tau}={d\psi^\mu\over d\tau}+ {\Gamma^\mu}_{\nu\rho}\psi^\nu
{dq^\rho\over d\tau}
\ \ \ \ \ \ \ {D\psi^*_\mu\over d\tau}={d\psi^*_\mu\over d\tau}
- {\Gamma^\nu}_{\mu\rho}\psi^*_\nu {dq^\rho\over d\tau}
\end{equation}
where ${\Gamma^\mu}_{\nu\rho}$ is the usual Christoffel connection on the
configuration space. The Riemann curvature tensor is defined as
\begin{equation}
R_{\mu\nu\rho\sigma} \equiv G_{\mu\lambda} (
{\Gamma^\lambda}_{\nu\sigma,\rho}-{\Gamma^\lambda}_{\nu\rho,\sigma}
+{\Gamma^\lambda}_{\tau\rho}{\Gamma^\tau}_{\nu\sigma} -
{\Gamma^\lambda}_{\tau\sigma}{\Gamma^\tau}_{\nu\rho}).
\end{equation}

This model can be quantised using either the canonical formalism or path
integrals.
We will briefly discuss the former. In the canonically quantised theory,
observables are
represented by operators which act on a multi-component wave function. The
different
components of the wave-function correspond to solutions with different fermion
number.

The wave function can be represented as a linear combination of differential
forms
on the configuration space \cite{Witten-deRham}. In this representation, the
operators
corresponding to the variables $\psi^\mu$ and
$\psi^*_\mu$ act on a differential form $\omega$ according
to the rule
\begin{equation}
\widehat\psi^\mu\omega = dq^\mu\wedge\omega,\ \ \ \
\widehat\psi^*_\mu\omega=-\hbar i_\mu \omega
\end{equation}
where $i_\mu\omega$ denotes the interior product of $\omega$ with
the tangent vector ${\partial\over\partial q^\mu}$.

The supersymmetry generators can be obtained from the action by the Noether
procedure. With the form of the Lagrangian given above, the supersymmetry
generators are represented by the operators
\begin{equation}
\widehat Q =  -\hbar e^{\Phi/\hbar} \d e^{-\Phi/\hbar},\ \ \ \ \
\widehat Q^* = \hbar e^{-\Phi/\hbar} \delt  e^{\Phi/\hbar},\ \ \ \
\end{equation}
where $\d$ is the usual exterior derivative, and $\delt$ is its
adjoint with respect to the standard inner product for differential forms,
$\langle \omega|\eta\rangle = \int \omega\wedge {}^*\eta.$

It will be useful to consider how these operators are affected by a
canonical transfomation of the classical theory. In particular, if the
Lagrangian
$L$ is augmented or diminished by the $\tau$-derivative of $\Phi
+\frac12 \psi^*_\mu\psi^\mu$,
\begin{equation}
L\mapsto L_\pm = L \pm {d\over d\tau} ( \Phi +\frac12 \psi^*_\mu\psi^\mu)
\label{eq:b_terms}
\end{equation}
then we obtain new representations of the quantum theory. These particular
choices of
the boundary action are needed below (eq(\ref{eq:ferm_int})) in order to ensure
the
existence of Nicolai maps \cite{L1}.

The wave functions $\omega_\pm$ in the new representations are related to the
wave function $\omega$ of the original representation by the transformation
\begin{equation}
\omega\mapsto \omega_\pm =e^{\mp \Phi/ \hbar}\omega.
\end{equation}
For the sake of notational simplicity, we have omitted the operator
$\frac12 \widehat\psi^*_\mu\widehat\psi^\mu/\hbar$ from the exponent.
It is easily seen that this omission affects only the relative normalisations
of the different components of the wave function.

The supersymmetry generators transform in a corresponding manner. Going to the
``$+$"
representation, one has
\begin{eqnarray}
\widehat Q\mapsto \widehat Q_+& =
e^{- \Phi/\hbar } \widehat Q
e^{+ \Phi/\hbar} & = -\hbar \d\ ,   \\
\widehat Q^* \mapsto \widehat Q^*_+ &
= e^{- \Phi/\hbar } \widehat Q^*  e^{\Phi/\hbar}
& = \hbar e^{-2\Phi/\hbar} \delt  e^{2\Phi/\hbar}
\end{eqnarray}
while going to the ``$-$" representation gives
\begin{eqnarray}
\widehat Q\mapsto \widehat Q_-& =
e^{+ \Phi/\hbar }   \widehat Q   e^{- \Phi/\hbar}&
= -\hbar e^{+2\Phi/\hbar} \d  e^{-2\Phi/\hbar},   \\
\widehat Q^* \mapsto \widehat Q^*_- &
= e^{+\Phi/\hbar }          \widehat Q^*       e^{-\Phi/\hbar}&
= \hbar \delt.
\end{eqnarray}

Whichever representation is used, the Hamiltonian operator is related to
the supersymmetry generators by the anticommutation relation
\begin{equation}
\widehat Q_\pm \widehat Q^*_\pm  + \widehat Q^*_\pm \widehat Q_\pm  =
-2\widehat H_\pm
\end{equation}
and the Euclidean-time evolution of the wave function is governed by the
Schr\"odinger equation
\begin{equation}
-\hbar {\partial{\omega_\pm}\over\partial\tau}= \widehat H_\pm \omega_\pm.
\label{eq:Schrodinger}
\end{equation}

\bigbreak

\leftline{\mainh Initial Conditions for Nicolai Maps.}\nobreak
\smallskip\nobreak
In the path-integral quantisation of the non-linear $\sigma$-model, every
Euclidean
history of the variables $q,\psi,\psi^*$ is assigned a definite weight.
A purely bosonic theory can be obtained by integrating out the fermionic
variables $\psi,\psi^*$. This integration yields a factor $J$ which,
in general, depends on the configuration of the bosonic variables
$q^\mu(\tau)$.

As shown by Nicolai, the supersymmetry of the underlying model ensures that
the factor $J$ may be interpreted as the Jacobian of a transformation from a
set of
non-interacting variables $\xi^a(\tau)$ to the variables $q^\mu(\tau)$
appearing in
the bosonic functional integral; this transformation is known as a Nicolai map.
In the case of Euclidean supersymmetric quantum mechanics, the Nicolai map
takes the form of a first-order stochastic diferential equation, in which the
variables
$\xi^a$ can be intepreted as uncorrelated Gaussian noise \cite{Gr3}.

To ensure that the Nicolai map is one-to-one, it is necesary to impose some
kind of boundary conditions on the bosonic variables $q^\mu$; for a stochastic
interpretation, the most natural choice is to specify the values of these
variables
at some initial time $\tau_0$.

However, Nicolai's theorem will not work unless the boundary conditions are
invariant
under either one or other of the supersymmetry generators. To ensure the
existence of
a Nicolai map, it is therefore necessary to supplement the initial conditions
on $q^\mu$
with Dirichlet initial conditions on either the variables $\psi^\mu$ or else
their
conjugates $\psi^*_\mu$ at $\tau_0$ \cite{L1}. Note that different choices of
initial conditions will result in different Nicolai maps.

It is useful to interpret these initial conditions in terms of the operators of
the canonical formalism.
Setting the initial values of $q^\mu$ means that, at time $\tau_0$, the wave
function
$\omega$ is in an eigenstate of the operator $\widehat q^\mu(\tau_0)$. Hence
$\omega(\tau_0)$ vanishes except at the specified value of $q^\mu$.

We have a choice of fermionic initial conditions. Imposing Dirichlet
conditions on $\psi^\mu$ means the wave function $\omega$ must satisfy
\begin{equation}
0 =\widehat\psi^\mu\omega=dq^\mu\wedge\omega \ \ \ \ \hbox{at } \tau_0  \ \ \
{\mu=1,\ldots n}
\label{eq:IC1}
\end{equation}
and consequently $\omega$ must be purely of degree $n$ at time $\tau_0$, where
$n$ is
the dimensionality of the configuration space.  It turns out that the
Hamiltonian operator commutes with the fermion number operator
$\widehat\psi^\mu\widehat\psi^*_\mu$ which yields the degree of the form, and
so
$\omega$ remains an $n$-form for all $\tau$.

The alternative choice is to impose Dirichlet initial conditions on the
variables
$\psi^*_\mu$ at $\tau_0$. When the wave function is represented by differential
forms,
these conditions are written
\begin{equation}
0 =\widehat\psi^*_\mu\omega=
-\hbar i_\mu\omega \ \ \ \ \hbox{at } \tau_0  \ \ \ {\mu=1,\ldots n}.
\label{eq:IC2}
\end{equation}
In this case, therefore, $\omega$ must be a pure 0-form at time $\tau_0$.
Again,
this property is preserved for all time $\tau$ since the Hamiltonian respects
fermion number.

The initial conditions necessary for the existence of a Nicolai map therefore
require that all but two of the components of the wave function vanish
identically. In other words, these two components alone can be generated from
Nicolai maps.
\bigbreak

\leftline{\mainh Construction of Nicolai Maps for Quantum Cosmology.}\nobreak
\smallskip\nobreak
Nicolai maps have been constructed for the supersymmetric non-linear
$\sigma$-models
described above, and quantum cosmological models can be treated as a special
case
of these. As remarked earlier, it suffices to consider globally supersymmetric
descriptions.

The first step in the construction of a Nicolai map is to obtain a purely
bosonic theory
by integration of the fermions subject to the appropriate initial conditions.
It is important here to use a form of the action which is invariant under the
same supersymmetry transformations as the initial conditions; if the
initial conditions (\ref{eq:IC1}) are imposed on $\psi^\mu$, then the $L_+$
form
of the Lagrangian must be used, while if the conditions (\ref{eq:IC2}) are
imposed
then $L_-$ must be used. Note that the Lagrangian must have one of the forms
specified in (\ref{eq:b_terms}) if there is to exist a Nicolai map \cite{Nic}.

Integration of the fermions then yields the following expression for the weight
of
a path $q^\mu(\tau)$ in configuration space;
\begin{eqnarray}
{\cal P}[q(\tau)]&= & \int d[\psi^*,\psi]
\exp -{1\over\hbar}
\left\{\int_{\tau_0}^{\infty}( L_\pm[q,\psi^*,\psi]\right\} \label{eq:ferm_int}
\\
  &=& J[q] \exp -{1\over\hbar} \left
\{\int_{t_0}^\infty
\frac12 ({ dq^\mu\over d\tau} \pm G^{\mu\nu}\Phi_{,\nu})^2 d\tau  \right\}
\end{eqnarray}
where the functional $J[q]$ depends on the Ricci curvature scalar of
the configuration space
according to the rule \cite{GRCH}
\begin{equation}
J[q]= \exp \pm {1\over 2\hbar}\int_{\tau_0}^\infty ( G^{\mu\nu}\Phi_{;\mu\nu} -
\frac14 R) d\tau.
\end{equation}

In fact the functional $J[q]$ is the Jacobian for the transformation
$\xi^a(\tau)\mapsto q^\mu(\tau)$ defined by the differential equation
\begin{equation}
{\Delta q^\mu\over d\tau}  = \mp G^{\mu\nu}(q) \Phi_{,\nu}  + e^\mu_a(q)\cdot
\xi^a
\label{eq:SDE}
\end{equation}
where $e^i_a(q)$ is a vielbein field on the configuration space with the
property that
\begin{equation}
G_{\mu\nu} e^\mu_a e^\nu_b =\eta_{ab}=\hbox{constant}.
\end{equation}
The differentials $\Delta q^\mu=dq^\mu + g^{\nu\rho} \Gamma^\mu_{\nu\rho}
d\tau$ and
$e^\mu_a(q)\cdot \xi^a d\tau$ are
defined so that they transform covariantly in the It\^o calculus \cite{Gr3}.
It follows by a change of variable that the weight for a given history
$\xi^a(\tau)$ of the new variable is
\begin{equation}
{\cal P}[\xi(\tau)] = \exp -{1\over 2\hbar}
\left\{\int_{\tau_0}^{\infty} \eta_{ab}\xi^a\xi^b d\tau \right\}
\label{eq:PI}
\end{equation}
so that $\xi^a(\tau)$, for positive definite $\eta_{ab}$, can be interpreted as
an
uncorrelated Gaussian noise process.

At this point we must acknowledge a conceptual difficulty arising from
this result; namely, that the minisuperspace metric $G_{\mu\nu}$ and
consequently
the ``noise'' metric  $\eta_{ab}$ both have a Lorentzian signature due to the
fact that the scale factor for the Universe is always a time-like variable.
Consequently, expression (\ref{eq:PI}) for the weight of the
path $\xi(\tau)$ is unbounded and can only be integrated by first rotating
$\xi^0$ to the imaginary axis. In eq.(\ref{eq:PI}) this rotation turns $\xi^0$
into noise with a positive intensity (like the other components) and changes
$G^{\mu\nu}$, which can be represented as
$G^{\mu\nu}=e^\mu_a e^\nu_b \langle \xi^a\xi^b\rangle d\tau$, to
$$\tilde G^{\mu\nu} = e_a^\mu e_b^\nu \delta^{ab}.$$
In fact this approach provides a definite prescription for dealing with the
problem of
unbounded action, which arises in different guises in all attempts to quantise
Einstein gravity.

The stochastic differential equation (\ref{eq:SDE}) is the Nicolai map for
the model, subject to the correponding set of initial conditions (\ref{eq:IC1})
or (\ref{eq:IC2}). After rotating to imaginary $\xi^0$, this is a Langevin
equation
for the particle whose motion in minisuperspace represents the Euclidean time
evolution
of the Universe. The associated Fokker-Planck equation takes the form
\begin{equation}
{\partial P  \over \partial \tau}=
\frac12 [\tilde G^{\mu\nu} (\hbar P_{,\nu} \pm 2\Phi_{,\nu}P) ]_{;\mu}
\label{eq:FP}
\end{equation}
where $P$ is positive. Moreover, the integral of $P(q,\tau)$ over the
minisuperspace
is conserved with respect to Euclidean time. These features justify referring
to
$P$ as the probability distribution function for the Universe.

Eq. \ref{eq:FP} can also be derived directly from the canonically quantised
theory
described earlier. Recall that initial conditions on $\psi^\mu$ imply that the
wave function is a $n$-form for all $\tau$, and consequently
$\widehat Q_+\omega_+$ vanishes in the ``$+$" representation.
It follows from (\ref{eq:Schrodinger}) that
\begin{equation}
{\partial\omega_+\over\partial\tau}= {1\over 2 \hbar}
\widehat Q_+ \widehat Q^*_+ \omega_+  =
-\frac12 \d (\hbar\delt \omega_+ + 2i_A\omega_+)
\end{equation}
where $i_A \omega_+$ stands for the interior product of the $n$-form
$\omega_+$ with the tangent vector $G^{\mu\nu}\Phi_{,\nu}{\partial\over\partial
q^\mu}$.
The Hodge star of $\omega_+$, denoted ${}^*\omega_+$, is then a pure scalar and
satisfies
\begin{equation}
{\partial ({}^*\omega_+) \over\partial\tau} =
- \frac12 \delt  \left[ \hbar\d ({}^*\omega_+) +
2\d\Phi\wedge({}^*\omega_+)\right] .
\label{eq:om1}
\end{equation}

Alternatively if one chooses the initial condition on $\psi^*_\mu$
then, using the ``$-$" repesentation, we know that $\widehat Q^*\, \omega_-$
vanishes
for all $\tau$ and therefore
\begin{equation}
{\partial\omega_-\over\partial\tau}= {1\over 2 \hbar}
\widehat Q^*_-\widehat Q_- \omega_-  =
-\frac12 \delt  \left[ \hbar\d  \omega_- - 2\d \Phi\wedge \omega_- \right]
\label{eq:om2}
\end{equation}
Identifying $P$ with either $\omega_-$ or ${}^*\omega_+$ then
reproduces eq. (\ref{eq:FP}), provided that $G^{\mu\nu}$ and $\delt$ are
analytically
continued according to the prescription defined earlier.

It is important to note that the wave function only satisfies a
conservation equation such as (\ref{eq:om1}) or (\ref{eq:om2}) if we use
the $\omega_+$ or $\omega_-$ representation; there is no conservation
equation in the canonically-related $\omega$ representation.

We recall that the original motive for considering the
Euclidean version of the theory was to enable us to find the
ground state for quantum cosmology, which is given by the
$\tau\mapsto\infty$ limit. As $\tilde G^{\mu\nu}$ is positive definite,
$P(q,\tau)$ approaches a stationary distribution of
the form
\begin{equation}
P_0(q)= A  \exp \pm 2\Phi(q)/\hbar. \label{eq:stat_dist}
\end{equation}
provided that this distribution is normalisable with the invariant measure
$$\left(Det (\tilde G_{\mu\nu}) \right)^{1/2}d^nq = |Det (G_{\mu\nu})|^{1/2}d^n
q.$$
Note that the distribution $P_0(q)$ is a static solution of eq.(\ref{eq:FP})
even if we
use the original minisuperspace metric $G^{\mu\nu}$ instead of its
positive-definite
analytic continuation $\tilde G^{\mu\nu}$.

Let us close by giving two examples of static solutions by considering
the case of Bianchi type IX. It turns out that there are in this case two
different
supersymmetric Lagrangians (\ref{eq:Lagrangian}) compatible with the same
Lagrangian in the classical limit \cite{BGr}. The two superpotentials are
\begin{equation}
\Phi={1\over 6}\left( e^{\beta^1} +e^{\beta^2} +e^{\beta^3} \right)
-2\hbar (\beta^1+ \beta^2+ \beta^3) \label{eq:potl1}
\end{equation}
and
\begin{equation}
\tilde\Phi={1\over 6}\left( e^{\beta^1} +e^{\beta^2} +e^{\beta^3}
-2e^{\beta^1+\beta^2} -2e^{\beta^2+\beta^3} -2e^{\beta^3+\beta^1} \right)
-2\hbar (\beta^1+ \beta^2+ \beta^3) \label{eq:potl2}
\end{equation}
where the variables $\beta^i$ parametrise the metric by
\begin{equation}
ds^2=-N(t)^2 + {1\over 6\pi} \sum_{i=1}^3 e^{2\beta^i} \omega^i \omega^i
\end{equation}
with the lapse function $N(t)$ and the basis 1-forms $\omega^i$  satisfying
$$d\omega^1= \omega^2\wedge \omega^2  \ \ \ \ \ \ \hbox{and cyclic
permutations.}$$
The terms proportional to $\hbar$ in eqs (\ref{eq:potl1}),(\ref{eq:potl2}) are
quantum
corrections whose given explicit form is obtained only if the larger $N=4$
supersymmetry suggested by dimensional reduction of $N=1$ supergravity is
imposed
\cite{Gr4}.
In the present context of $N=2$ supersymmetry the form of these terms could be
left
arbitrary, reflecting an operator-ordering ambiguity. It is interesting,
however, that
with $\Phi$ as given, the probability distribution $\exp -2\Phi/\hbar$ is
normalisable.
This state can be interpreted as a wormhole state.

The distribution $\exp -2\tilde\Phi/\hbar$, on the other hand, is not
normalisable.
However this problem is not as serious as it first seems. Defining a scale
factor
$$a=\exp \left[(\beta^1+\beta^2+\beta^3)/3\right] $$
we see that, with $\beta^1-\beta^2$ and $\beta^1+\beta^2-2\beta^3$ fixed, the
normalisation integral converges as $a\to 0 $ when $\tilde\Phi$ has the form
given in
(\ref{eq:stat_dist}). Although the integrand diverges as $a\to\infty$,
this is not a serious concern as the role played by matter fields
(which we have neglected)
becomes important for large $a$ and eqs (\ref{eq:stat_dist}),(\ref{eq:potl2})
can no
longer be expected to provide a good approximation to the true wave function.

Interestingly, the probability distributions given above are precisely the same
as those
obtained recently by Bene and Graham \cite{BGr} using a different and more
conventional
approach, in which the wave function is intepreted as the square root of the
probability
density.

\bigbreak

\leftline{\mainh Discussion and Conclusion}\nobreak
\smallskip\nobreak

We have shown that the existence of an N=2 supersymmetric extension of the
minisuperspace Wheeler-DeWitt equations suggests a stochastic interpretation
of the resulting solutions. It appears likely that a stochastic interpretation
of
a similar kind will also apply to more general cosmological models with various
types of
extended supersymmetries (including full supergravity) and work to establish
this is in
progress. In such cases the existence of Nicolai maps will be assured only if
the
initial conditions we impose are invariant under some non-trivial subalgebra of
the full
supersymmetry algebra \cite{L1}. Assuming that this subalgebra includes the
generators
of Lorentz transformations, it follows that boundary conditions for Nicolai
maps
should be Lorentz invariant. In particular, this means that Dirichlet initial
conditions must be imposed on \it all \rm the components of either the
left-handed
or the right-handed spinors (though not on both, or the path integral would be
overdetermined). If chirality is preserved by the equations of motion,
these same components will vanish throughout the subsequent evolution. We can
therefore surmise that, in such cases, Nicolai maps will exist only for states
in which spinors of a certain chirality vanish identically.

For example, let us consider the case of pure $N=1$ supergravity without a
cosmological constant. (Our analysis will also apply to minisuperspace models
with $N=4$ supersymmetry obtained from this theory by dimensional reduction).
In this case, the chirality of the spinor fields is preserved by their
equations of motion.
Consequently, if we impose Dirichlet initial conditions on the left-handed
spinors
$\overline\psi^{A'}$ in order to obtain a state admitting a Nicolai map,
then the $\overline\psi^{A'}$ will vanish at all subsequent times.
The wave-function will then depend only on the right-handed spinors $\psi^A$
and the bosonic variables. However, the momenta conjugate to the $\psi^A$ are
linear combinations of the $\overline\psi^{A'}$, which vanish identically.
It follows that the wave function must in fact be independent of the $\psi^A$,
and therefore is simply a function of the bosonic variables.

We therefore conclude that any boundary conditions which permit Nicolai maps in
this model will give rise to states in one of the bosonic sectors. (There are
two
bosonic sectors, depending on whether we choose $\psi^A$ or
$\overline\psi^{A'}$ as the
canonical coordinates.These correspond to the empty and filled fermion states
discussed
earlier.)  Interestingly, these are precisely the sectors which are found to
admit
physical solutions in the Bianchi I case considered by  D'Eath, Hawking and
Obr\'egon \cite{DHO} and the Bianchi IX case \cite{Gr4,D2}
as well as the more general Bianchi class A models studied by Asano,
Tanimoto and Yoshino \cite{ATY}.

Note that the above argument doesn't work if we include matter or a
cosmological constant,
since then the chirality of spinors is not preserved by their equations of
motions.
In these cases, therefore, we would \it not \rm expect states in the bosonic
sectors
to admit Nicolai maps. But in the models of this kind which have been studied
so far
\cite{Gr2,BGr}, it turns out that there are in fact no physical states in these
sectors!

In view of these connections, it is tempting to speculate that the physical
states in
supersymmetric quantum cosmological models are precisely those which admit
Nicolai
maps. This would lend considerable support to the stochastic intepretation of
the
wave function outlined above. However, much further work is needed to clarify
this
question.

\bigskip
\bigskip
\bigbreak

\leftline{\bf Acknowledgements}
H.L. is grateful to the Universit\"at Gesamthochschule Essen, and R.G to the
University of Sydney, for hospitality during recent visits. This work was
supported
by the Deutsches Forschungsgemeinschaft through the Sonderforschungbereich 237
``Unordnung und grose Fluktuationen", and by a Sydney University Research
Grant.

\vfill
\end{document}